**Detector Backaction on the Self-Consistent Bound State in Quantum Point Contacts**

Y. Yoon[1], M.-G. Kang[1], T. Morimoto[2] (森本崇弘), L. Mourokh[3], N. Aoki[4] (青木伸之),
J. L. Reno[5], J. P. Bird[1], and Y. Ochiai[4] (落合勇一)

1: Department of Electrical Engineering, University at Buffalo, the State University of New York, Buffalo, NY 14260-1920, USA

2: Advanced Device Laboratory, RIKEN, 2-1 Hirosawa, Wako, Saitama 351-0198, Japan

3: Department of Physics, Queens College of CUNY, 65-30 Kissena Blvd., Flushing, NY 11367, USA

4: Graduate School of Advanced Integration Science, Chiba University, 1-33 Yayoi-cho, Inage-ku, Chiba 263-8522, Japan

5: CINT Science Department, Sandia National Laboratories, P.O. Box 5800, Albuquerque, NM 87185-1303

## Abstract

Bound-state (BS) formation in quantum point contacts (QPCs) may offer a convenient way to localize and probe single spins. In this letter, we investigate how such BSs are affected by monitoring them with a second QPC, which is coupled to the BS via wavefunction overlap. We show that this coupling leads to a unique detector backaction, in which the BS is weakened by increasing its proximity to the detector. We also show, however, that this interaction between the QPCs can be regulated at will, by using an additional gate to control their wavefunction overlap.

The manner in which a measuring apparatus affects a quantum system has long occupied a central place in philosophical discussions of quantum theory. Mesoscopic devices, such as quantum dots (QDs) and quantum point contacts (QPCs), are well suited to the study of this problem, since they may easily be configured so that their properties are affected by coupling them to other, similar devices. QDs [1] and QPCs [2] can be used, for example, as capacitively-coupled charge detectors, to count electrons on a nearby, but electrically isolated, QD. In a solid-state realization of the which-path experiment [3], this approach was used to study the decoherence in an Aharonov-Bohm ring, due to the *backaction* exerted on carriers in one of its arms by a capacitively-coupled QPC “detector”. Charge sensing with QPCs has also been used to readout the results of spin-sensitive manipulation of single electrons on QDs [4-6] and QD molecules [7-10], all important steps for the implementation of solid-state quantum computing. A general problem with the capacitive-sensing scheme, however, arises from the backaction of the detector, the shot-noise in whose current can induce undesirable transitions in the system under study [11,12].

Coupled mesoscopic structures have also been used to study spin transport in QPCs [13-18], whose carriers are thought to spontaneously spin polarize close to pinch-off [19]. In one interpretation, the QPC has been argued to function as a single-spin trap [20-22] that confines an electron to a bound state (BS), formed naturally by the Friedel oscillations [22,23] in the self-consistent potential profile of the QPC. Evidence of such spin binding has been found by studying the conductance of a (detector) QPC in close proximity to another (swept QPC) [13-15]. A resonance occurs in the detector conductance when the swept QPC pinches off and has been attributed [16,17] to an unusual Fano effect [24,25], arising from the wavefunction overlap between the BS and the detector. The idea is that a coherent tunnel correlation arises between the QPCs, with electrons continuously being swapped between them, when the BS is driven through the Fermi level by the swept-QPC gate voltage (Fig. 1(a)). The interference of these partial waves can then be shown [16] to give rise to a resonance in the detector conductance, just as is seen in experiment. The resonance develops Zeeman splitting in a magnetic field [14], consistent with occupation of the BS by a single, well-defined spin [17].

The use of coupled QPCs as an electrically-addressable single-spin system could be extremely important, particularly for efforts to implement solid-state quantum computing. Realizing such applications in practice, however, will require a proper understanding of how the bound spin on one QPC is affected by its coupling to the detector. Theoretically, at least, the resonance exhibited by the QPCs is thought to arise from a very different mechanism to capacitive charge sensing, involving instead single-particle interference due to the wavefunction overlap of the QPCs [16,17]. Such a mechanism has the potential to allow for spin readout with low detector decoherence, since it should avoid multi-electron shot-noise dephasing [5,8-10], and since it should also be possible to control the wavefunction overlap on-demand using appropriate gates [17]. To investigate these issues, in this letter we study how the BS on a QPC is modified by the proximity of its detector. We show that the detector response is indeed consistent with a Fano effect, arising from its wavefunction overlap with the BS, and also show how the interaction between the QPCs may be regulated by modulating the degree of this overlap. Most importantly, we find that the wave-mechanical coupling between the QPCs leads to an unusual backaction, in which increasing the proximity of the two QPCs dramatically suppresses the robustness of their resonance. This behavior is inconsistent with quantum transitions generated by detector shot-noise, and is attributed instead to a detector-induced weakening of the BS self-consistent confining potential.

We studied the multi-gate GaAs/AlGaAs (Sandia sample EA750) device of Ref. [14], whose two-dimensional electron gas (2DEG) was of density $2.3 \times 10^{11}$ cm$^{-2}$, mobility $4 \times 10^{6}$ cm$^{2}$/Vs, and mean free path 31 microns, at 4.2 K. The mean free path decreased to 4 microns by 77 K, but even here significantly exceeded the largest inter-QPC separation (~800 nm) used in experiment. The schematic of Fig. 1(b) labels the different gates and Ohmic contacts used in our experiments, which involved applying fixed voltage ($V_d$) to a pair of gates to form the detector QPC, while varying the gate voltage ($V_s$) used to form the swept QPC. The conductance ($G_s$) of this QPC was first measured as a function of $V_s$, while leaving the Ohmic contacts of the detector floating. After this, the detector conductance ($G_d$) was measured for the same range of $V_s$, but with the swept-QPC Ohmics floating. Previous work showed that $G_d$ exhibits a resonance as the swept-QPC pinches off [14,15], which is observed in QPCs with various configurations [13,14]. $G_s$ & $G_d$ were measured by low-frequency lock-in detection, with an excitation of 30 μV (unless stated otherwise) and at temperatures from 4.2 – 60 K.

We begin our discussion of the backaction by demonstrating that the resonant response of the detector is consistent with a mechanism involving wavefunction overlap between the QPCs, rather than Coulomb coupling. Fig. 1(c) shows the resonance obtained in the usual configuration, with the two QPCs coupled via their intervening 2DEG [14]. In Fig. 1(d), however, we repeat the same measurement, while using an additional gate ($G_5$) to cut off this coupling by fully depleting the 2DEG in its vicinity. The detector resonance is clearly quenched, which would not be expected if it were due to Coulomb coupling of the two QPCs. This quenching is consistent, however, with the interaction arising from the wavefunction overlap of the QPCs. (In contrast, the linear background on which the resonance is superimposed [14] is unaffected by cutting off the 2DEG coupling. This feature *is* therefore likely due to the direct electrostatic influence of $V_s$ on $G_d$.)

The coupling of the QPCs through their intervening 2DEG gives rise to a distinct *configuration* dependence to the detector resonance, as we illustrate in Fig. 2. Figs. 2(a) & 2(b) show the resonance obtained [14] for the largest possible QPC separation, while Figs. 2(e) & 2(f) show that for the smallest separation (~300 nm) achievable in our device. Figures 2(c) & 2(d) are for the configuration intermediate between these limits, and a comparison of all of the data shows that the detector-resonance evolves systematically with QPC separation. While the resonance has been attributed [16,17] to a Fano effect, this property is not immediately obvious in Figs. 2(a) & 2(b), whose resonances do not exhibit the strong asymmetry usual of the Fano effect [24,25]. In Figs. 2(e) & 2(f), however, the resonances clearly show the classic Fano form, with a deep minimum in immediate proximity to a local maximum. While prior work has suggested the existence of a BS in QPCs [26,27], our observation of a clear Fano effect due to this BS is the most direct evidence for its existence to date.

Figure 2 shows that the lineshape of the detector resonance is not unique to any single QPC, but rather is a property of the coupled-QPC configuration. Consider Figs. 2(d) & 2(e), in which the swept QPC is the same, but in which the distance to the detector is different. A clear Fano form is obtained with the swept QPC close to the detector (Fig. 2(e)). For increased separation (Fig. 2(d)), however, the Fano asymmetry is less pronounced. The systematic dependence on detector proximity is confirmed by the very similar form of the resonances in Figs. 2(c) & 2(d), and in Figs. 2(e) & 2(f), which utilize different gates but correspond to equivalent coupled-QPC configurations. It is furthermore confirmed in Fig. 3, which shows resonances obtained with different sets of gates used to implement equivalent multi-QPC configurations, corresponding to the minimum possible separation between the two QPCs. In all four cases, the detector resonance shows a very-similar Fano form to that in Figs. 2(e) & 2(f).

Quite generally, the symmetry of Fano resonances is related to the strength of the coupling between their resonant and non-resonant channels [24], being more asymmetric for increased coupling. Our results (Fig. 2), showing that the detector resonance becomes more asymmetric when the inter-QPC separation is reduced, thereby yielding stronger wavefunction overlap, are quite consistent with this. For further analysis, we fit the detector resonance to the Fano form [24,25], $G_d(\varepsilon) \propto (\varepsilon + q)^2/(\varepsilon^2 + 1)$, where $\varepsilon \equiv 2(V_g - V_o)/\Gamma$, and $V_o$ and $\Gamma$ are, respectively, the resonance position and width. $q$ is a parameter related to the strength of the coupling between the different channels and determines the symmetry of the resonance. Quite generally, with increasing coupling, the magnitude of $q$ decreases (from $q = \infty$ for no coupling) and the asymmetry of the resonance becomes more pronounced [24,25]. In our fitting, we account for the linear background to $G_d$ (Fig. 1(d)), which we have already noted is of separate origin to the interference effect that yields the detector resonance, and use $V_o$, $q$ and $\Gamma$ as parameters. The fits are shown in Figs. 2 & 3 and reproduce the experimental behavior. In spite of the multi-parameter nature of the fits, we emphasize that it is clear to even the naked eye in Fig. 2 that the Fano asymmetry is more pronounced when the QPCs are closer. Fig. 1(e) plots averaged values of $q$ for different configurations and shows a systematic decrease of $q$ as the QPC separation is reduced, indicating increased coupling of the resonant and non-resonant channels [24]. This is again consistent with a wavefunction-based interaction between the QPCs [16,17], and emphasizes that even the weakly-asymmetric peaks in Figs. 2(a) – 2(d) should be viewed as Fano resonances, albeit ones with weaker Fano coupling due to the larger QPC separation.

The non-capacitive nature of the inter-QPC interaction gives rise to an unusual detector backaction, revealed in studies of the temperature dependence of the resonance. For configurations similar to those in Figs. 2(a) & 2(b), this was previously found to persist to around 40 K, indicating a robust (~meV) confinement of the BS on the swept QPC [14]. While surprising for a coherent effect, the survival of the resonance (albeit strongly damped) to such high temperatures can likely be attributed to the proximity of the QPCs, which should allow even a small fraction of carriers entering the 2DEG from the detector to scatter coherently from the BS at higher temperatures. In Fig. 4, however, we demonstrate that bringing the QPCs in *too close* proximity can actually *suppress* the detector resonance. In this figure, we show that in configuration B the resonance is similarly robust to that found previously. In configuration A, however, for which the QPCs are in maximal proximity, the wash out of the resonance is dramatically suppressed to ~10 K. This suppression is not specific to any QPC, but is rather common to configurations in which the QPCs are in close proximity (~300 nm). The same wash-out behavior is obtained, for example, by reversing the role of swept and detector QPCs, or by using other sets of gates to implement the same configuration (as in Fig. 3). Since the resonance arises from the interference of partial waves that travel directly from the detector to the drain, via the 2DEG, and those that scatter from the BS (Fig. 1(a)) [16,17], the rapid wash out of the resonance in configuration A might be due to enhanced detector-induced dephasing when the QPC separation is reduced. The right inset of Fig. 4 shows the resonance in this configuration, however, with the detector excitation voltage (and current) increased by a factor of ten. While the increased current should enhance shot-noise induced dephasing, the wash-out of the Fano resonance is unaffected. We therefore conclude that shot noise is not responsible for the rapid wash out of the resonance when the QPC separation is reduced.

BS formation in QPCs has been argued to occur when the spin-dependent, self-consistent interactions among carriers create a potential well that localizes a single spin on the QPC [20-22]. The main result of this Letter is that the act of “observing” this BS with another QPC may weaken its confinement, in a manner that depends systematically on the proximity of the two QPCs. The weakening becomes pronounced when the QPC separation is reduced, and is accompanied by a growth in the asymmetry of the Fano resonance, which in itself indicates an enhancement of the wavefunction overlap between the QPCs [24]. These results suggest a unique form of detector backaction, in which the self-consistent potential that confines the BS is weakened by its direct wavefunction overlap with the detector. While detailed calculations are needed to clarify the nature of the interaction between the QPCs, we note that BS formation is expected as a consequence of the multiple scattering of electron waves from the bare poten-tial barrier associated with the QPC [20-22]. The presence of a second QPC in close proximity likely modifies this self-consistent scattering, although it is not clear how this might weaken the BS potential. One possibility is that, when the proximity of the QPCs is increased, the local potential variations that they generate overlap with each other, thereby modifying the properties of the BS. We have argued already [16] that the BS-induced scattering of electron waves that emanate from the detector is responsible for its resonant response. It is therefore possible that, as the wavefunction overlap is increased by decreasing the inter-QPC separation, this scattering itself weakens the BS confinement. Regardless of the actual mechanism, our experiment unambiguously demonstrates the sensitivity of the BS properties to the local environmental conditions, something that is quite consistent with the notion that BS formation is a self-consistent process. We have also seen that the wavefunction coupling of the QPCs can be cut off at will (Fig. 1(d)), and this control could possibly allow implementation of a spin readout scheme with low detector-induced decoherence, as previously proposed in Ref. [17].

This work was supported by the DoE (DE-FG03-01ER45920) and was performed, in part, at the Center for Integrated Nanotechnologies, a U.S. DoE Office of Basic Energy Sciences nanoscale science research center. Sandia National Laboratories is a multi-program laboratory operated by Sandia Corporation, a Lockheed-Martin Company, for the U.S. DoE (Contract No. DE-AC04-94AL85000).

## FIGURE CAPTIONS

**Fig. 1:** (a) Schematic of the resonance when electrons tunnel back and forth between a BS on the swept QPC (foreground) and the detector (background). (b) Device schematic. $G_1$ – $G_8$ are gates on the semiconductor surface and Ohmic contacts are numbered 1 – 8. (c) Variation of $G_d$ & $G_s$ (line with open circles) at 4.2 K. $G_d$ is measured passing current ($I$) from 1 to 4, and measuring voltage ($V$) between 2 & 3. $G_s$ uses 8 & 5 for $I$ and 7 & 6 for $V$. Circle in schematic indicates where BS is formed. (d) Same as (c) but with gate $G_5$ biased to pinch off the region between $G_5$ & $G_8$. (e) Fano factor for different inter-QPC separations. Error bars denote the spread of $q$ values for different configurations.

**Fig. 2:** Resonance in $G_d$ (open symbols, solid line through symbols denotes fit to Fano form) and variation of $G_s$ for different QPC configurations at 4.2 K. The corresponding Fano factor ($q$) is shown. Circle in schematics indicates where BS is formed. (a) $G_d$: $I$ – 1 & 4, $V$: 2 & 3, $G_s$: $I$ – 8 & 5, $V$: 7 & 6. (b) $G_d$: $I$ – 8 & 5, $V$: 7 & 6, $G_s$: $I$ – 1 & 4, $V$: 2 & 3. (c) $G_d$: $I$ – 1 & 4, $V$: 2 & 3, $G_s$: $I$ – 1 & 8, $V$: 2 & 7. (d) $G_d$: $I$ – 8 & 5, $V$: 7 & 6, $G_s$: $I$ – 1 & 8, $V$: 2 & 7. (e) $G_d$: $I$ – 1 & 4, $V$: 2 & 3, $G_s$: $I$ – 1 & 8, $V$: 2 & 7. (f) $G_d$: $I$ – 8 & 5, $V$: 7 & 6, $G_s$: $I$ – 1 & 8, $V$: 2 & 7.

**Fig. 3:** Resonance in $G_d$ (open symbols, solid line through symbols denotes fit to Fano form) and variation of $G_s$ for equivalent QPC configurations at 4.2 K. The corresponding Fano factor ($q$) is shown. Circle in schematics indicates where BS is formed. (a) $G_d$: $I$ – 1 & 8, $V$: 2 & 7, $G_s$: $I$ – 1 & 4, $V$: 2 & 3. (b) $G_d$: $I$ – 1 & 8, $V$: 2 & 7, $G_s$: $I$ – 8 & 5, $V$: 7 & 6. (c) $G_d$: $I$ – 4 & 5, $V$: 3 & 6, $G_s$: $I$ – 8 & 5, $V$: 7 & 6. (d) $G_d$: $I$ – 8 & 5, $V$: 7 & 6, $G_s$: $I$ – 4 & 5, $V$: 3 & 6.

**Fig. 4:** Main panel: temperature dependence of detector resonance amplitude for the gate configurations (A & B) indicated on the upper right-hand side of the figure. Insets show detector resonance for different temperatures for the two configurations. Configuration A: $G_d$: $I$ – 1 & 4, $V$: 2

& 3 $G_s$: $I$ – 1 & 8, $V$: 2 & 7. Configuration B: $G_d$: $I$ – 1 & 4, $V$: 2 & 3 $G_s$: $I$ – 1 & 8, $V$: 2 & 7. Results for configuration A are shown for a measurement excitation of 30 μV (solid line) and 300 μV (line with open symbols).

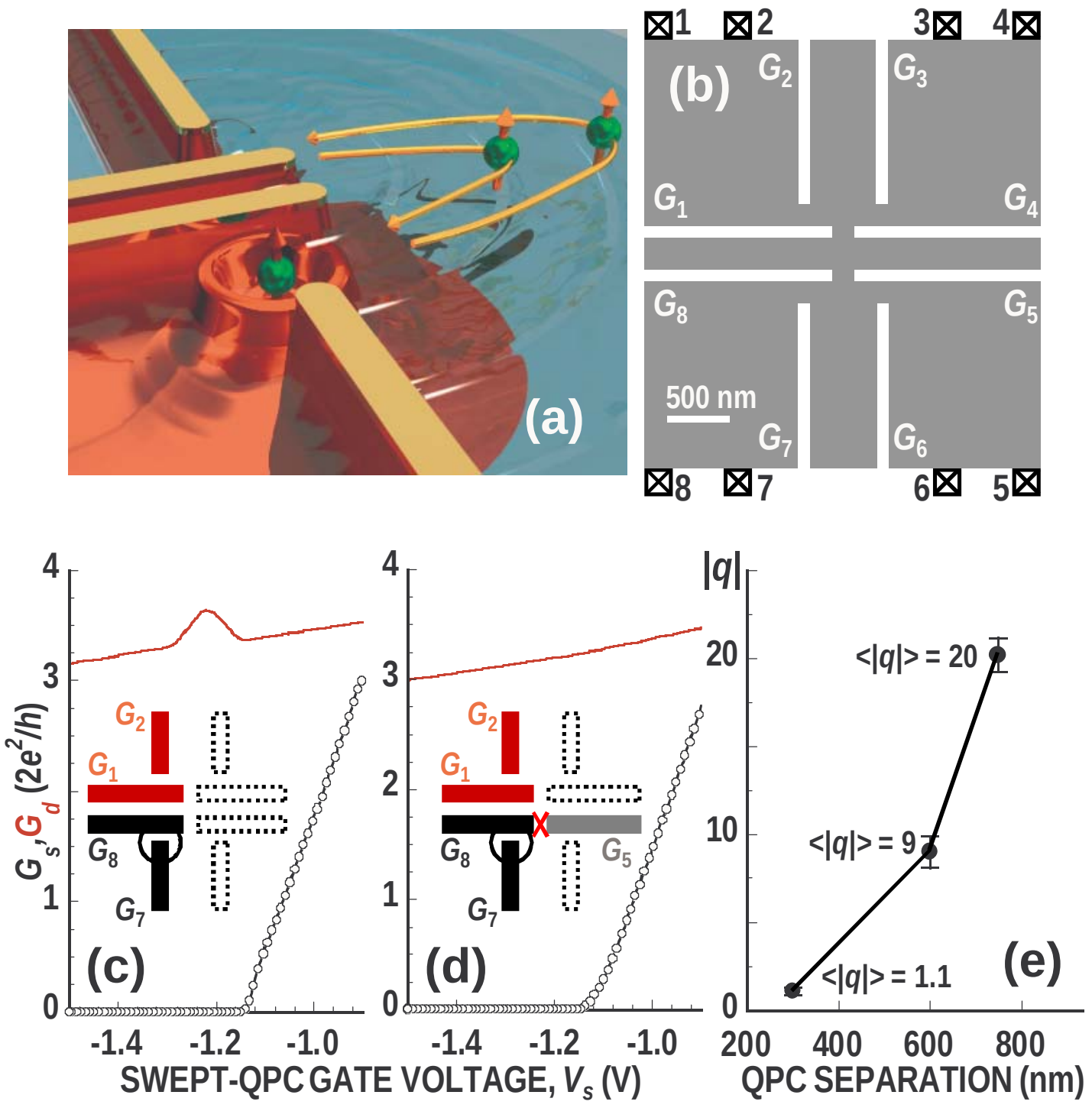
(a)
(b)
1
2
3
4
8
7
6
5
G2
G3
G1
G4
G8
G5
G7
G6
500 nm
(c)
(d)
(e)
Gs, Gd (2e²/h)
SWEPT-QPC GATE VOLTAGE, Vs (V)
|q|
<|q|> = 20
<|q|> = 9
<|q|> = 1.1
QPC SEPARATION (nm)

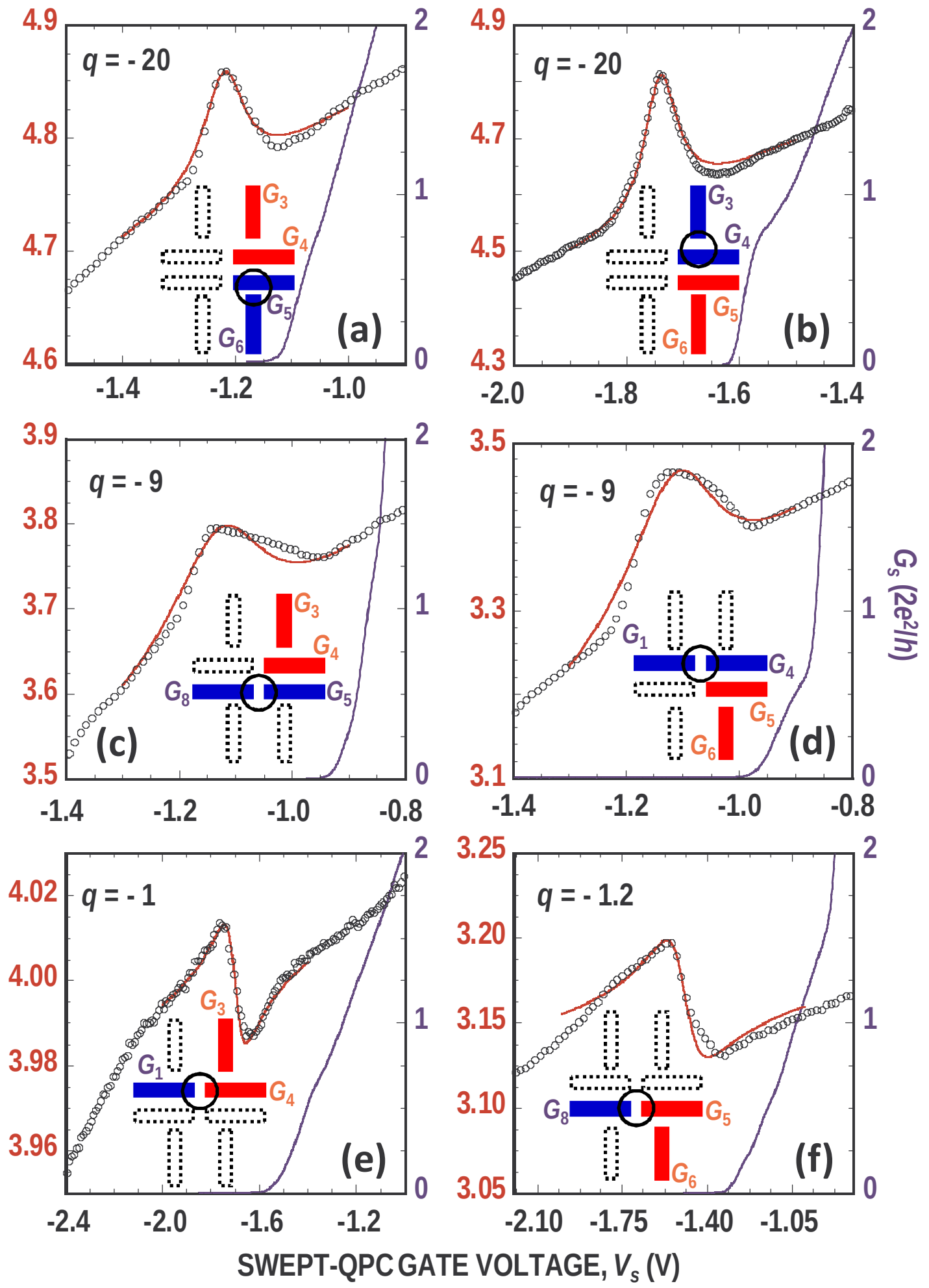

q = - 20
(a)
q = - 20
(b)
q = - 9
(c)
q = - 9
(d)
q = - 1
(e)
q = - 1.2
(f)
G3
G4
G5
G6
G1
G8
Gd (2e2/h)
Gs (2e2/h)
SWEPT-QPC GATE VOLTAGE, Vs (V)

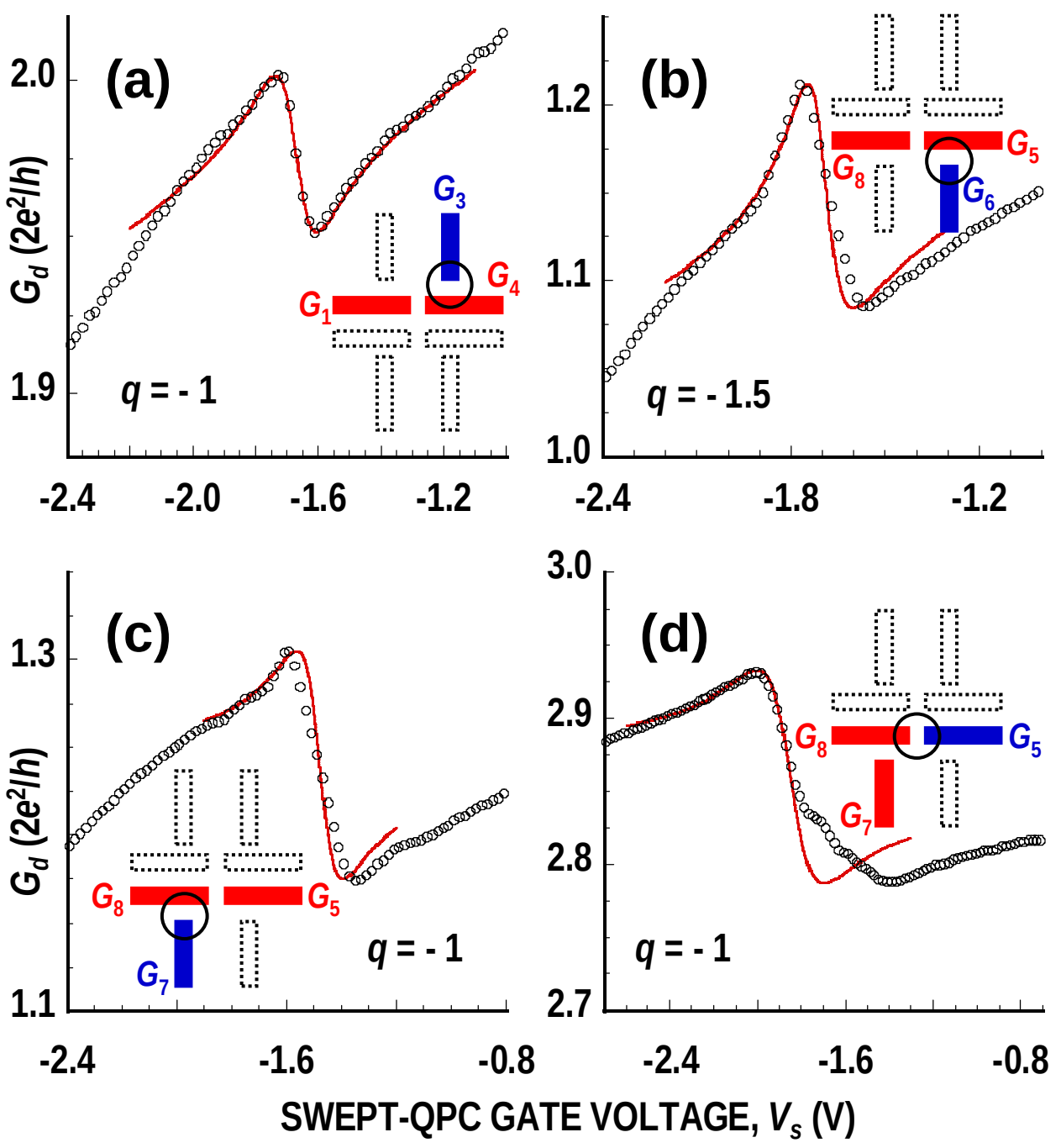

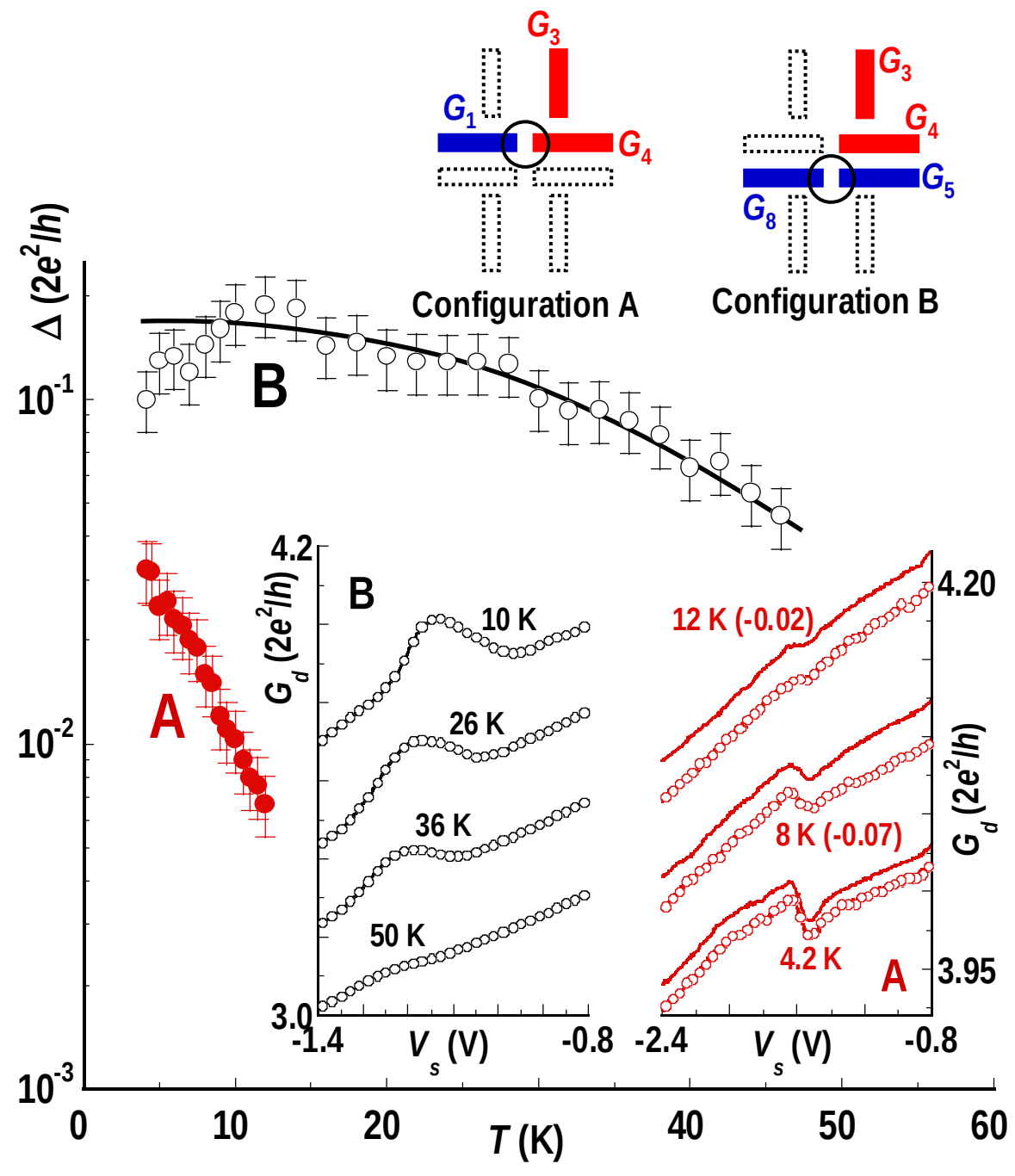